# Lower bound on neutrino mass and possible CP violation in neutrino oscillations

Rohit Verma

*Rayat Institute of Engineering and Information Technology, Ropar 140001, India*
*e-mail address:* rohitverma@live.com

The phenomenology of the most general lepton mass matrices obtained through weak basis transformations has been discussed. Using a hierarchical parametrization of these mass matrices, the exact relations for lepton mixing angles have been obtained assuming normal neutrino mass hierarchy and natural structures of lepton mass matrices. The recent three neutrino oscillation data together with the condition of naturalness on Dirac lepton mass matrices provide a lower bound on the lightest neutrino mass of $m_{\nu 1} > 3$ meV along with a non-vanishing Dirac CP violating phase.

## I. INTRODUCTION

The recent measurement of large $\theta_{13}$ [1] has not only provided a big impetus to the search for CP violation in the lepton sector, but also deepened the mystery of the origin of neutrino masses and mixing angles which now appear to be significant different from the masses and the flavor mixing patterns of quarks. As the mixing angles are related to the corresponding mass matrices, therefore attempts to unravel the mystery of fermion mass generation and mixings become more complicated, especially if the mass matrices for quarks and leptons are to be considered on the same footing [2].

In case neutrinos are considered to be Dirac particles, they can acquire masses exactly in the same way as quarks and charged leptons do in the standard model. In this context, it has been shown that the highly-suppressed Yukawa couplings for Dirac neutrinos can naturally be achieved in the models with extra spacial dimensions [3] or through radiative mechanisms [4]. In the present work, we simply assume neutrinos to be Dirac particles and perform a systematic study of the most general lepton mass matrices.

One of the several phenomenological approaches used to extract clues for the formulation of fermion mass matrices is the 'texture zero' approach initiated by Weinberg and Fritzsch [5]. Subsequently, several texture based lepton mass matrices have been investigated in the literature [5].

Recently [7], it has been shown that some sets of these zeros have, by themselves, no physical meaning, since these can be obtained starting from arbitrary fermion mass matrices by making appropriate unitary transformations also called weak basis (WB) transformations. It was also observed [7][8] that using the freedom of such transformations, it is possible to obtain Hermitian fermion mass matrices involving a 'maximum' of three phenomenological texture zeros. Any 'additional' texture zero is supposed to have physical implications.

Among the several possibilities of texture based lepton mass matrices, all texture six zero possibilities appear to be ruled out [9][10] while texture five [9][11] and four [12] zero lepton mass matrices are able to explain the lepton mixing for the case of Dirac neutrinos with normal mass hierarchy (NH). However all such mass matrices involve more than three texture zeros and may not be considered to be 'general' in view of the WB transformations.

The purpose of the present paper, on one hand, is to probe the neutrino masses and possible CP violation in neutrino oscillations using exact relations for the lepton mixing angles wherein the effect of the neutrino mass hierarchy on the mixing angles is clearly evident, while on the other hand, implications of the most general lepton mass matrices on observed neutrino oscillations have been investigated. Noting that the possibility of Dirac neutrinos may still be allowed by the experiments [13] and that the inverted neutrino mass hierarchy appears to be ruled out for lepton mass matrices [9][10][11] involving a greater number of texture zeros, the analysis has been limited to the study of *most general texture based Hermitian* lepton mass matrices obtained through weak basis transformations and have assumed neutrinos to be Dirac particles exhibiting NH.

## II. WB TRANSFORMATIONS

In the WB approach [7][8], one usually considers the mass matrices for charged leptons $M_e$ and neutrinos $M_\nu$, respectively, as

$$M_e = D_e, \quad M_\nu = V D_\nu V^\dagger = \begin{pmatrix} e_\nu & |a_\nu|e^{i\alpha_\nu} & |f_\nu|e^{i\omega_\nu} \\ |a_\nu|e^{-i\alpha_\nu} & d_\nu & |b_\nu|e^{i\beta_\nu} \\ |f_\nu|e^{-i\omega_\nu} & |b_\nu|e^{-i\beta_\nu} & c_\nu \end{pmatrix}. \quad (1)$$

Here $D_e = \text{diag}(m_e, m_\mu, m_\tau)$, $D_\nu = \text{diag}(m_{\nu 1}, m_{\nu 2}, m_{\nu 3})$ and V is the Pontecorvo-Maki-Nakagawa-Sakata (PMNS) matrix [14],[15].

It has been shown [16] that for the quark sector, the observed hierarchy among the quark masses as well as the Cabibbo-Kobayashi-Maskawa quark mixing matrix [15] elements gets naturally translated onto the structure of the corresponding quark mass matrices i.e.

$$e < (|a|, |f|) < d < |b| < c. \quad (2)$$

Such hierarchical mass matrices have been referred to in the literature as *natural mass matrices* [17]. In principle, an exact diagonalization of the mass matrix given in eqn. (1) is not always possible. In this context, one can apply a WB transformation [7][8] U on the mass matrices $M_e$ and $M_\nu$, such that

$$M_e \to M'_e = U M_e U^\dagger, \quad M_\nu \to M'_\nu = U M_\nu U^\dagger. \quad (3)$$

The two representations $(M_e, M_\nu)$ and $(M'_e, M'_\nu)$ are physically equivalent and result in the same PMNS matrix. As discussed in [8], there is a possible choice of U such that

$$(M'_e)_{13,31} = (M'_\nu)_{13,31} = (M'_\nu)_{11} = 0, \quad (4)$$

or

$$(M'_\nu)_{13,31} = (M'_e)_{13,31} = (M'_e)_{11} = 0, \quad (5)$$

with non-vanishing other elements. This necessitates an investigation of the physical implications of the above general lepton mass matrices $(M'_e, M'_\nu)$ for neutrino mixing phenomenology, especially if the condition of naturalness in eqn. (2) is imposed on these.

## III. WB TEXTURE DIAGONALIZATION

Noting that for the WB textures of lepton mass matrices in eqns. (4) and (5), one of the lepton mass matrices is a Fritzsch-like texture two zero type [8] whereas the other has the following form

$$M'_L = \begin{pmatrix} e_L & |a_L|e^{i\alpha_L} & 0 \\ |a_L|e^{-i\alpha_L} & d_L & |b_L|e^{i\beta_L} \\ 0 & |b_L|e^{-i\beta_L} & c_L \end{pmatrix}, \quad L = e, \nu. \quad (6)$$

It is observed [16][18] that for $|a_L|$ and $|b_L|$ to be real,

$$m_1 > e_L > -m_2,$$
$$(m_3 - m_2 - e_L) > d_L > (m_1 - m_2 - e_L). \quad (7)$$

where the indices 1, 2, 3 ≡ e, μ, τ for $M_e$ and ν1, ν2 and ν3 for $M_\nu$. The small positive values of $e_L$ are consistent with the condition of naturalness. The diagonalizing matrix $O_e$ for $M'_e$ defined through $D_e = O_e^T P_e M'_e P_e^\dagger O_e$ with $P_e = \text{diag}(e^{-i\alpha_e}, 1, e^{i\beta_e})$, may be expressed as [16] $O_e =$

$$\begin{pmatrix} 1 & \sqrt{\dfrac{m_{e\mu}(1-\xi_e)}{(1+m_{\mu\tau})}} & \sqrt{\dfrac{m_{e\tau}m_{\mu\tau}(\zeta_e+m_{\mu\tau})(1-\xi_e)}{(1+m_{\mu\tau})}} \\ \sqrt{\dfrac{m_{e\mu}(1-\xi_e)}{(1+\zeta_e)}} & -\sqrt{\dfrac{1}{(1+\zeta_e)(1+m_{\mu\tau})}} & \sqrt{\dfrac{(\zeta_e+m_{\mu\tau})}{(1+\zeta_e)(1+m_{\mu\tau})}} \\ -\sqrt{\dfrac{m_{e\mu}(\zeta_e+m_{\mu\tau})(1-\xi_e)}{(1+\zeta_e)}} & \sqrt{\dfrac{(\zeta_e+m_{\mu\tau})}{(1+\zeta_e)(1+m_{\mu\tau})}} & \sqrt{\dfrac{1}{(1+\zeta_e)(1+m_{\mu\tau})}} \end{pmatrix},$$

(8)

where $m_e \ll m_\mu$ and $m_e \ll m_\tau$ have been used for the charged lepton masses. The free parameters $\xi_e$ and $\zeta_e$ represent the hierarchy characterizing parameters for the mass matrix $M'_e$ and are defined as $\xi_e = e_e/m_e$, $\zeta_e = d_e/c_e$ while $m_{e\mu} = m_e/m_\mu$, $m_{e\tau} = m_e/m_\tau$ along with $m_{\mu\tau} = m_\mu/m_\tau$ have been considered for simplicity.

However, for the Dirac neutrino mass matrix $M'_\nu$, the diagonalizing transformation $O_\nu$ defined through $D_\nu = O_\nu^T P_\nu M'_\nu P_\nu^\dagger O_\nu$, is given by $O_\nu =$

$$\begin{pmatrix} \sqrt{\dfrac{(1+\xi_\nu m_{\nu 12})}{(1+m_{\nu 12})}} & \sqrt{\dfrac{m_{\nu 12}(1-\xi_\nu)}{(1+m_{\nu 12})(1+m_{\nu 23})}} & \kappa\sqrt{\dfrac{m_{\nu 13}m_{\nu 23}(m_{\nu 23}+\zeta_\nu)(1+\xi_\nu m_{\nu 12})}{(1-\zeta_\nu)(1+m_{\nu 23})}} \\ \sqrt{\dfrac{m_{\nu 12}(1-\xi_\nu)(1-\zeta_\nu)}{(1+m_{\nu 12})}} & -\sqrt{\dfrac{(1+\xi_\nu m_{\nu 12})(1-\zeta_\nu)}{(1+m_{\nu 12})(1+m_{\nu 23})}} & \sqrt{\dfrac{(m_{\nu 23}+\zeta_\nu)}{(1+m_{\nu 23})}} \\ -\sqrt{\dfrac{m_{\nu 12}(m_{\nu 23}+\zeta_\nu)}{(1+m_{\nu 12})}} & \sqrt{\dfrac{(1+\xi_\nu m_{\nu 12})(m_{\nu 23}+\zeta_\nu)}{(1+m_{\nu 12})(1+m_{\nu 23})}} & \sqrt{\dfrac{(1-\zeta_\nu)}{(1+m_{\nu 23})}} \end{pmatrix},$$

(9)

where $P_\nu = \text{diag}(e^{-i\alpha_\nu}, 1, e^{i\beta_\nu})$, $\kappa = \sqrt{(1-\xi_\nu)/(1-m_{\nu 13})}$ and the free parameters $\xi_\nu = e_\nu/m_{\nu 1}$, $\zeta_\nu = d_\nu/m_{\nu 3}$ characterize the hierarchy that is exhibited by the elements of the Dirac neutrino mass matrix, while $m_{\nu 12} = m_{\nu 1}/m_{\nu 2}$, $m_{\nu 13} = m_{\nu 1}/m_{\nu 3}$ and $m_{\nu 23} = m_{\nu 2}/m_{\nu 3}$ have again been considered for simplicity. The corresponding lepton mass matrices assume the natural form of eqn. (2) if $(\zeta_e, \zeta_\nu, \xi_e, \xi_\nu) < 1$ and Dirac neutrinos exhibit NH i.e. $m_{\nu 1} < m_{\nu 2} < m_{\nu 3}$.

## IV. CONSTRUCTING PMNS MATRIX

One can now easily compute the PMNS matrix through $V = O_e^\dagger P_e P_\nu^\dagger O_\nu$. In general,

$$V_{i\sigma} = O_{1i}^e O_{1\sigma}^\nu e^{-i\phi_1} + O_{2i}^e O_{2\sigma}^\nu + O_{3i}^e O_{3\sigma}^\nu e^{i\phi_2} \quad (10)$$

where the phases $\phi_1 = \alpha_e - \alpha_\nu$ and $\phi_2 = \beta_e - \beta_\nu$ are also free parameters.

### A. CASE I

For the WB representation $(M'_e, M'_\nu)$ given in eqn. (4), the three lepton mixing angles may be quite accurately expressed as

$$s_{12} = \sqrt{\dfrac{m_{\nu 12}}{(1+m_{\nu 12})(1+m_{\nu 23})}}, \quad (11)$$

$$s_{13} = \left| \sqrt{\dfrac{m_{\nu 13}m_{\nu 23}(m_{\nu 23}+\zeta_\nu)}{(1-\zeta_\nu)(1-m_{\nu 13})(1+m_{\nu 23})}} e^{-i\phi_1} - \sqrt{\dfrac{m_{e\mu}(1-\xi_e)}{(1+\zeta_e)(1+m_{\nu 23})}} \left( \sqrt{(m_{\nu 23}+\zeta_\nu)} - \sqrt{(\zeta_e+m_{\mu\tau})(1-\zeta_\nu)} e^{i\phi_2} \right) \right|, \quad (12)$$

$$s_{23} = \left| \sqrt{\dfrac{1}{(1+\zeta_e)(1+m_{\mu\tau})(1+m_{\nu 23})}} \left( \sqrt{(m_{\nu 23}+\zeta_\nu)} - \sqrt{(\zeta_e+m_{\mu\tau})(1-\zeta_\nu)} e^{i\phi_2} \right) \right|, \quad (13)$$

where only the leading order term (first) as well as the next to leading order terms have been retained in the expressions. It is observed that the above relations hold good within an error of less than a percent. It is noteworthy that the mixing angle $s_{12}$ depends only on the neutrino mass ratios $m_{v12}$ and $m_{v23}$. Likewise, it is also observed that the mixing angle $s_{23}$ is independent of $\xi_e$. This is easy to interpret as $\xi_e$ does not invoke any mixing among the second and the third generations of leptons. As a result, it should be interesting to investigate the implications of $\xi_e$, if any, for $s_{13}$ as well as those of $\zeta_e$ and $\zeta_v$ for $s_{13}$ and $s_{23}$.

### B. CASE II

For the WB representation $(M'_e, M'_v)$ given in eqn. (5), the mixing angles can be expressed as

$$s_{12} = \sqrt{\frac{m_{v12}(1-\xi_v)}{(1+m_{v12})(1+m_{v23})}}, \quad (14)$$

$$s_{13} = \left| \sqrt{\frac{m_{v13}m_{v23}(m_{v23}+\zeta_v)(1+\xi_v m_{v12})(1-\xi_v)}{(1-\zeta_v)(1-m_{v13})(1+m_{v23})}} e^{-i\phi_1} - \sqrt{\frac{m_{e\mu}}{(1+\zeta_e)(1+m_{v23})}} \left( \sqrt{(m_{v23}+\zeta_v)} - \sqrt{(\zeta_e+m_{\mu\tau})(1-\zeta_v)} e^{i\phi_2} \right) \right|, \quad (15)$$

$$s_{23} = \left| \sqrt{\frac{1}{(1+\zeta_e)(1+m_{\mu\tau})(1+m_{v23})}} \left( \sqrt{(m_{v23}+\zeta_v)} - \sqrt{(\zeta_e+m_{\mu\tau})(1-\zeta_v)} e^{i\phi_2} \right) \right|. \quad (16)$$

One observes that the mixing angle $s_{12}$ depends on the neutrino mass ratios $m_{v12}$, $m_{v23}$ and the parameter $\xi_v$. As expected, the mixing angle $s_{23}$ is still independent of $\xi_v$. As a result, it should be interesting to investigate the implications of $\xi_v$, if any, for $s_{12}$ and $s_{13}$ as well as those of $\zeta_e$ and $\zeta_v$ for $s_{13}$ and $s_{23}$.

### V. INPUTS

The following 1σ C.L. values for the various three neutrino mixing parameters [19] have been used as inputs for the analysis, i.e.

$$\delta m^2 = (7.32 - 7.80) \times 10^{-5} \text{ GeV}^2,$$
$$\Delta m^2 = (2.33 - 2.49) \times 10^{-3} \text{ GeV}^2,$$
$$\sin^2 \theta_{12} = 0.29 - 0.33,$$
$$\sin^2 \theta_{13} = 0.022 - 0.027,$$
$$\sin^2 \theta_{23} = 0.37 - 0.41, \quad (17)$$

where the neutrino mass square differences are defined as $\delta m^2 = m_{v2}^2 - m_{v1}^2$ and $\Delta m^2 = m_{v3}^2 - (m_{v1}^2 + m_{v2}^2)/2$ for NH [19]. The eqns. (11) and (12) imply a clear constraint on the neutrino mass ratios $m_{v12}$ and $m_{v13}$ and hence on the neutrino mass $m_{v1}$ through the neutrino oscillation parameters $s_{12}$ and $s_{13}$, since

$$m_{v12} = \sqrt{\frac{m_{v1}^2}{m_{v1}^2 + \delta m^2}} \quad \text{and} \quad m_{v13} = \sqrt{\frac{m_{v1}^2}{m_{v1}^2 + \Delta m^2 + (\delta m^2)/2}}. \quad (18)$$

More explicitly, the lightest neutrino mass $m_{v1}$ along with the parameters $\zeta_e, \zeta_v, \xi_e, \xi_v$ and phases $\phi_1$ and $\phi_2$ are taken to be free parameters. In addition, the condition of naturalness has been imposed on the lepton mass matrices through the constraints $(\zeta_e, \zeta_v, \xi_e, \xi_v) < 1$ and NH has been assumed for the Dirac neutrinos, consistent with the condition of natural mass matrices. Furthermore, in the absence of any clues for CP violation in the lepton sector, the phases $\phi_1$ and $\phi_2$ have been given full variations.

### VI. RESULTS

### A. CASE I

It is observed that the complete 1σ range of all the neutrino oscillation parameters given in eqn. (17) can be reconstructed by the relations (11), (12) and (13). Interestingly, the condition of naturalness $\zeta_v < 1$ on the Dirac neutrino matrix limits the phenomenologically allowed range of $\xi_e$ from $0 < \xi_e < 1$ to $0 < \xi_e < 0.86$ as shown in the FIG. 1. This may be attributed to the second term on R.H.S. of eqn. (12) which contributes a little less than the leading order term for estimating $s_{13}$. One can check the due to the presence of $\sqrt{(1-\xi_e)(m_{v23}+\zeta_v)}$ in this term, the large values of $\xi_e$ can couple only with large values of $\zeta_v$ to regenerate $s_{13}$ within the observed experimental range.

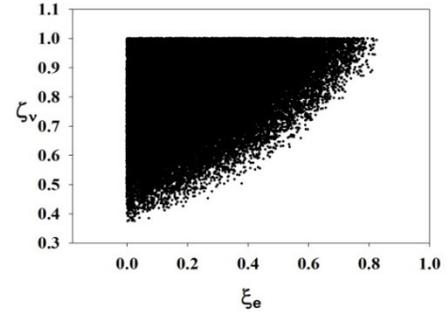

FIG. 1: The allowed parameter spaces of the free parameters $\xi_e$ and $\zeta_v$ in case I.

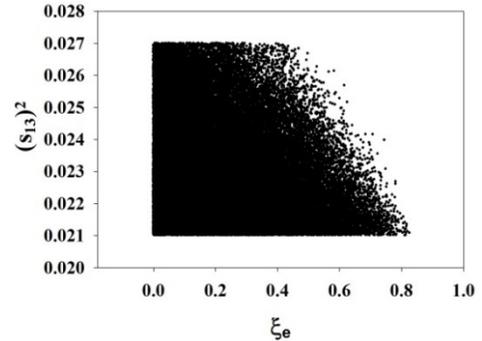

FIG. 2: The variation of $(s_{13})^2$ vs. $\xi_e$ in case I.

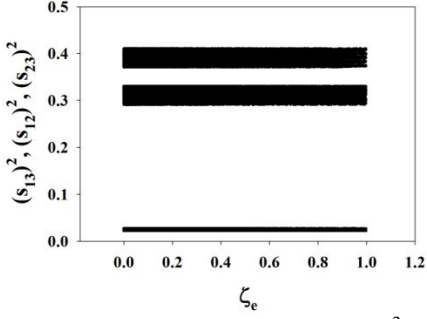

FIG. 3: Plot showing no dependance of $(s_{13})^2$, $(s_{12})^2$ and $(s_{23})^2$ on $\zeta_e$ in case I.

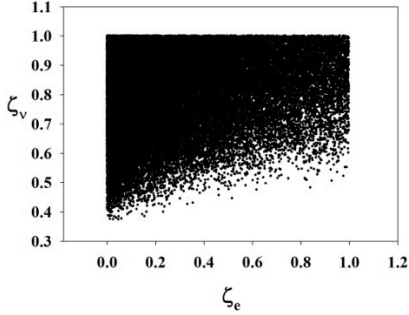

FIG. 4: The variation in the parameter $\zeta_v$ vs. the parameter $\zeta_e$ in case I.

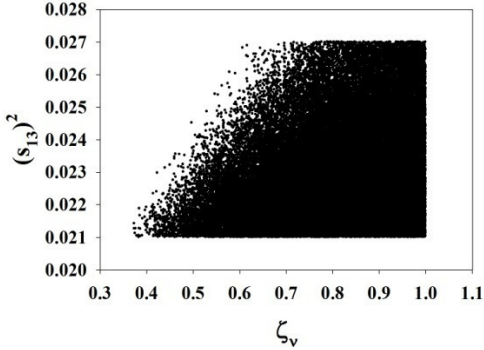

FIG. 5: Plot showing dependence of $(s_{13})^2$ on $\zeta_v$ in case I.

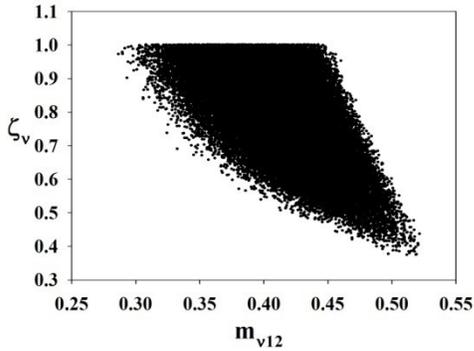

FIG. 6: Plot depicting that the lower bound on $m_{v12}$ is a direct consequence of $\xi_v < 1$ in case I.

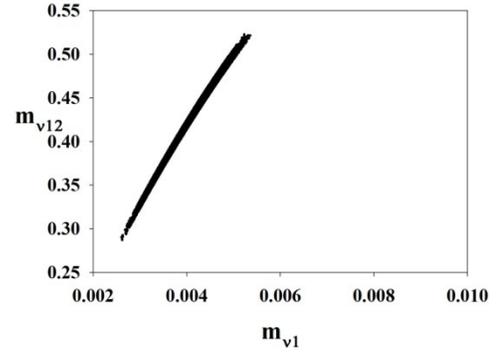

FIG. 7: Plot depicting the variation of $m_{v12}$ as a function of $m_{v1}$.

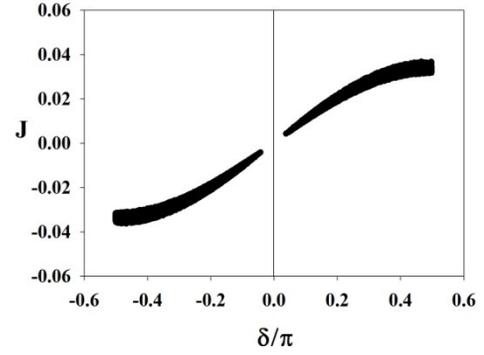

FIG. 8: The variation of Jarlksog rephasing invariant parameter J with the Dirac phase $\delta/\pi$ in case I.

Since the upper limit of $\zeta_v$ is constrained by the condition of naturalness $\zeta_v < 1$, as a result, one is not able to reconstruct $s_{13}$ for large values of $\xi_e$ as shown in FIG. 2. As expected, it is observed that that the mixing angles $s_{23}$ and $s_{12}$ are independent of the parameter $\xi_e$. As a result, it may be concluded that the allowed values for $\xi_e$ including $\xi_e = 0$ are able to reconstruct the entire 1σ range of $s_{13}$, $s_{23}$ and $s_{12}$ indicating that the parameter $\xi_e$ has no physical implications for lepton mixings. It is also observed that the parameter space for $\zeta_e$ allowed by the condition of naturalness does not seem to play a vital role in fixing the mixing angles, as depicted in FIG. 3. This is attributed to the terms $\sqrt{(1+\zeta_e)}$ in the denominators of eqns. (12) and (13) wherein the effects for non-vanishing $\zeta_e$ can be compensated by the freedom of the parameter space available to $\zeta_v$, as shown in FIG. 4, reinforcing that, like the parameter $\xi_e$, the parameter $\zeta_e$ has no physical implications for lepton mixings, and may also be considered as redundant. However, the same is not observed to be true for $\zeta_v$ as seen in FIG. 5, which shows that the values of $\zeta_v <$ 0.34 are not able to reproduce the mixing angle $s_{13}$. This may be understood using eqn. (12) which predicts that for very small values of $\zeta_v$, the leading order term should not able to regenerate the experimentally allowed $s_{13}$ implying that $\zeta_v = 0$ bears physical implications for neutrino oscillation phenomenology.

Interestingly, it is also observed that the lower bound of $m_{v12} > 0.287$ is a direct consequence of condition of naturalness on $M'_v$ given by $\zeta_v < 1$, and illustrated in FIG. 6. Since the ratio $m_{v12}$ is directly proportional to the lightest neutrino mass $m_{v1}$, depicted in eqn. (18) and FIG. 7, the corresponding lower bound on the lightest neutrino mass is found to be $m_{v1} > 2.59$ meV with $m_{v1} = (2.59 - 5.40)$ meV, $m_{v2} = (8.96 - 10.30)$ meV and $m_{v3} = (48.7 - 50.5)$ meV. The results that $s_{12}$ and $s_{13}$ appear to invoke strong constraints on the neutrino mass ratios $m_{v12}$ and $m_{v13}$ <u>are</u> in good agreement with the conclusions of ref. [9].

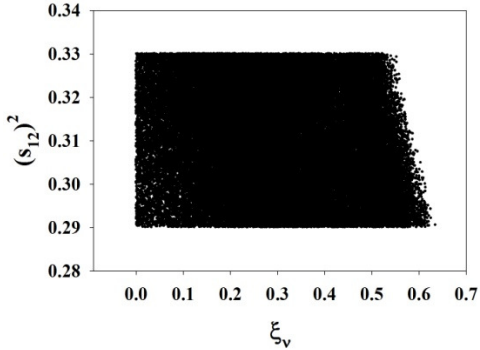

FIG. 9: Variation of $(s_{12})^2$ with $\xi_v$ for case II

It is noteworthy that the recent three neutrino oscillation data forbids 'both' the lepton mass matrices to be real. Using FIG. 8, it may be concluded that the recent three neutrino oscillation data requires the CP violating Dirac phase $\delta \neq 0$ requiring that at least one of the phases $\phi_1$ and $\phi_2$ must be non-vanishing.

## B. CASE II

Interestingly, the complete $1\sigma$ range of all the neutrino oscillation parameters given in eqn. (17) can also be regenerated by the relations (14), (15) and (16). However, from the eqn. (14) it is noticed that due to the term $\sqrt{(1-\xi_v)}$, the phenomenologically allowed range of $\xi_v$ is limited from $0 < \xi_v < 1$ to $0 < \xi_v < 0.64$ in order to regenerate the experimentally measured $s_{12}$, as illustrated in the FIG. 9.

It is also observed that the entire $1\sigma$ range of $s_{23}$ and $s_{13}$ can be reconstructed by the allowed parameter space for $\xi_v$. This is obvious from eqn. (16) for $s_{23}$. However, in the case of eqn. (15), the effect of increase in $\xi_v$ on $\sqrt{(1-\xi_v)}$ in the leading order term is compensated by a corresponding increase in $\sqrt{(1+\xi_v m_{v12})}$ and hence the $1\sigma$ range of mixing angle $s_{13}$ is regenerated in totality by the allowed parameter space for $\xi_v$. As a result, we notice that the non-vanishing values for $\xi_v$ including $\xi_v = 0$ are able to reconstruct the observed $1\sigma$ range of $s_{13}$, $s_{23}$ and $s_{12}$ reinforcing that the parameter $\xi_v$ also has no physical implications for lepton mixing. It is also noticed that, like $\xi_v$, $\zeta_e$ has no physical implications for the mixing angles, with similar reasoning as applicable in case I, indicating the redundancy of these parameters in the corresponding mass matrices.

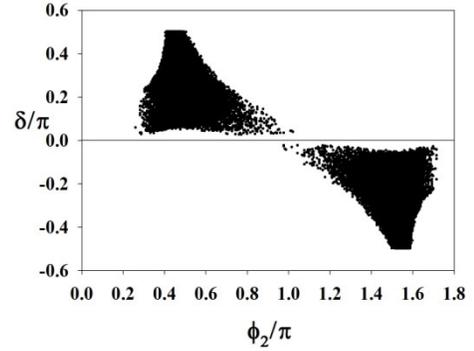

FIG. 10: Plot showing $\delta/\pi$ vs. $\phi_2/\pi$ for case II.

However, the values of the parameter $\zeta_v < 0.05$ are not able to reproduce the mixing angle $s_{13}$ implying that $\zeta_v = 0$ does have physical implications for lepton mixing. Interestingly, it is again observed that the condition of naturalness $\zeta_v < 1$ on the neutrino matrix provides a lower bound on $m_{v1} > 2.68$ meV. For this case, we obtain $m_{v1} = (2.68 - 10.0)$ meV, $m_{v2} = (9.0 - 13.3)$ meV and $m_{v3} = (48.7 - 51.3)$ meV. Furthermore, the real lepton mass matrices continue to be forbidden in this case as well, with no observed data points corresponding to $\delta = 0$ as shown in FIG. 10.

## VII. CONCLUSIONS

The analysis shows that, even though the neutrino mixing pattern is significantly different from the quarks mixing pattern, yet it can also be described by 'natural' mass matrices. Using a hierarchical parameterization for the lepton mass matrices as well as the condition of naturalness i.e. $(\zeta_e, \zeta_v, \xi_e, \xi_v) < 1$, we have been able to illustrate the effect of lepton mass hierarchies on lepton mixing through exact relations wherein the three lepton mixing angles are completely expressible in terms of the lepton mass ratios, the hierarchy characterizing parameters and the phases $\phi_1$ and $\phi_2$. Assuming NH for Dirac neutrinos, it has been clearly shown that for the recent three neutrino oscillation data, the most general texture three zero lepton mass matrices of eqns. (4) & (5), obtained through WB transformations, are physically equivalent to texture five zero Fritzsch-like Hermitian lepton mass matrices with $\xi_e = 0$, $\xi_\nu = 0$, $\zeta_e = 0$ and $\zeta_\nu \neq 0$, when the condition of naturalness is imposed on these. The corresponding mixing angles may be quite accurately expressed by Eqs. (11)-(13) with $\xi_e = 0$ and $\zeta_e = 0$ or using Eqs. (14)-(16) with $\xi_\nu = 0$ and $\zeta_e = 0$. It appears that the phenomenological difference between the three zero (WB choice) and Fritzsch-like Hermitian five zero (two assumptions added) textures is physically insignificant.

The naturalness condition on $M_\nu$ for these texture five zero lepton matrices is observed to provide a lower bound on $m_{\nu 1} > 3.03$ meV with $m_{\nu 1} = (3.03 - 5.45)$ meV, $m_{\nu 2} = (9.14 - 10.4)$ meV and $m_{\nu 3} = (48.8 - 50.5)$ meV.

These values for neutrino masses appear to favor standard leptogenesis as the mechanism to produce the Baryon Asymmetry of the Universe [20]. It is noteworthy that real structures of lepton mass matrices with $(\phi_1 = \phi_2 = 0)$ are forbidden by the current three neutrino oscillation data, indicating a possible CP violation in the lepton sector with $\delta \neq 0$ and leave the doors open for the possibility of Dirac nature of neutrinos with normal mass hierarchy.

## ACKNOWLEDGEMENT

The author would like to thank the Director, Rayat Institute of Engineering and Information Technology, for providing the necessary working facilities.